# Long-term evolution of regulatory DNA sequences.
# Part 1: Simulations on global, biophysically-realistic genotype-phenotype maps.

Elia Mascolo[1], Réka Borbély[1], Santiago Herrera-Álvarez[2], Calin C Guet[1], Justin Crocker[2], and Gašper Tkačik[1]

[1]Institute of Science and Technology Austria, Am Campus 1, AT-3400 Klosterneuburg, Austria

[2]Developmental Biology Unit, European Molecular Biology Laboratory, DE-69117 Heidelberg, Germany

**Abstract.** Promoters and enhancers are cis-regulatory elements (CREs), DNA sequences that bind transcription factor (TF) proteins to up- or down-regulate target genes. Decades-long efforts yielded TF-DNA interaction models that predict how strongly an individual TF binds arbitrary DNA sequences and how individual binding events on the CRE combine to affect gene expression. These insights can be synthesized into a global, biophysically-realistic, and quantitative genotype-phenotype (GP) map for gene regulation, a "holy grail" for the application of evolutionary theory. A global map provides a rare opportunity to simulate long-term evolution of regulatory sequences and pose several fundamental questions: How long does it take to evolve CREs *de novo*? How many non-trivial regulatory functions exist in sequence space? How connected are they? For which regulatory architecture is CRE evolution most rapid and evolvable? In this article, the first of a two-part series, we briefly review the pertinent modeling and simulation efforts for a unique system that enables close, quantitative, and mechanistic links between biophysics, as well as systems, synthetic, and evolutionary biology.

Much is understood about the power of evolutionary optimization. Assuming we know how the genetic program ("genotype") maps into observable organismal properties ("phenotype") and, subsequently, into fitness [1], a mathematically rigorous body of theory can predict evolutionary trajectories, diversity of outcomes, and the rates of adaptation in various population genetic regimes [2]. Thus, the complete knowledge of a genotype-phenotype-fitness map implies a full knowledge of evolutionary change. In practice, however, our knowledge is remarkably incomplete: we know neither all the relevant phenotypes nor how they map into fitness. Furthermore, the map typically depends on the environment in complex or unknown ways. Arguably the biggest hurdle, however, is the jump from genotypes to phenotypes: the curse of dimensionality generally precludes us from assigning a phenotype to each of the $4^L$ possible genotypes even in sequence spaces of modest length, e.g., for $L \gtrsim 10$ bp.

We typically circumvent this curse of dimensionality by bitter sacrifice. On the theory front, we study toy model genotype-phenotype-fitness maps or "fitness landscapes", where every possible genotype is mathematically assigned a fitness value. This exhaustive assignment to all $4^L$ possible genotypes defines what we call a "global map". On such generic and stylized landscapes (often going by inventive or cryptic names such as the house-of-cards, Mount Fuji, NK, pairwise- or globally-epistatic landscapes, etc.), one can study global, arbitrarily long evolutionary trajectories. We refer to such approaches as addressing "long-term evolution" if: evolution can start from any sequence (even fully random sequences, enabling the simulation of *de novo* evolution); it can proceed indefinitely, with no imposed limit to how many mutations can be accumulated. However, the use of idealized maps sacrifices biological realism and a quantitative match to any real dataset: at best, we might hope to capture generic properties of evolutionary dynamics; at worst, our toy model map may have missed some essential topological feature of the real map, putting even qualitative predictions into doubt.

On the empirical front, in contrast, experiments are typically designed to measure putative fitness proxies using massive, controlled mutational libraries. For example, fluorescence has been measured for thousands of GFP protein mutants [3,4], yielding landscapes that enable guided design of new GFP variants. Similarly, for transcriptional regulation, constitutive or regulated gene expression has been measured at scale using massively parallel assays [5,6]. In both cases, experiments provide a quantitative, system-specific map. The sacrifice here is the map's global nature: even the largest libraries probe only a vanishingly small fraction of the $4^L$ possible genotypes, confined to a local mutational neighborhood of the wild-type. We refer to such restricted assignments as "local maps". This limitation is not fatal when considering point mutations over sufficiently short timescales. We refer to this regime as "short-term evolution", characterized by two constraints: evolution starts from one or a few related sequences (typically the observed wild-type); it is limited to only a handful of mutations ($\lesssim 10$) that explore only the local mutational neighborhood around the initial sequences (the mutants for which experimental data is available). However, such approaches preclude more general predictions over arbitrarily long timescales (during which evolution may explore farther regions of genotype space), or modeling *de novo* evolution from non-functional sequences. Recent successes of AlphaFold [7] indicate that even this barrier might be overcome, if the protein structure can be predictively linked to its quantitative function and used as a fitness proxy. At the moment, however, raw maps derived empirically by massively parallel assays do not extrapolate well across the entire sequence space, leaving fundamental questions about long-term evolution largely out of reach.

GP maps that are both quantitative and global, and therefore uniquely suited to simulate long-term evolution, are few and far between. The established ones share a common theme: the phenotype depends on "molecular recognition", the propensity of two molecules to interact with a strength set by their sequence-dependent structure. Because physicochemical laws strongly constrain the form of the underlying interaction rules, the complexity of the GP map is drastically reduced. The prime example is the genetic code, where serial molecular recognition by tRNAs renders the GP map so simple that the phenotype (the protein sequence) can be decoded with a simple lookup table that assigns 21 messages (20 amino acids and a *stop* message) to the $4^3=64$ codewords of 3 bp (codons). The success and centrality of this reconstruction arguably diverted attention from comparable efforts to decode non-protein-coding DNA.

For other, more complex GP maps, existing massively parallel experiments can be used to quantitatively calibrate theory-derived molecular interaction rules, which in turn generalize from measured genotypes to the entire sequence space. The first landmark success in this direction has been the prediction of secondary RNA structures from their sequence by the "Vienna school" [8]. Subsequent efforts focused on molecular recognition in antigen-antibody interactions, emphasizing their specificity for pathogens and simultaneous avoidance of self-interactions, essential for healthy functioning of immune systems [9]. Last but not least, TF-DNA interactions that we focus on in this review series have also been a case-in-point for physically-informed quantitative GP maps that enable (semi-)realistic simulations of long-term CRE evolution.

The scope of this review series is purposefully narrow: To evaluate our progress in simulating and theoretically understanding long-term regulatory sequence evolution. This goal closely interacts with several disciplines. From the population genetics perspective, it presents a unique opportunity to apply and ultimately test population genetics – a mature mathematical theory – on evolutionary questions that are inaccessible for most other biological systems. For example, while we cannot expect the theory to answer "How long does it take to evolve an eye?" quantitatively, we may have a hope at "How long does it take to evolve a developmental enhancer?" From the comparative genomics perspective, regulatory sequence evolution is essential as it is thought to drive a large fraction of evolutionary adaptations in metazoans [10], including primates [11]. From the systems and synthetic biology perspective, predicting gene expression from regulatory sequence is one of the fields' defining questions, even though it is mostly pursued with minimal reference to evolutionary implications. Yet such predictive models pursued in systems biology are, quite literally, the GP maps for gene regulation, when viewed through the evolutionary lens and assuming, importantly, that the molecular mechanisms of regulatory sequence readout do not change on CRE evolution timescales. Recent years have seen an explosion of successful predictive models that combine large-scale experiments, biophysical constraints, and deep learning; we point the reader to several reviews on the topic [12–14].

To set the stage, Section I provides a rudimentary account of the essential building blocks of regulatory GP maps, with minimal detail and references. Section II summarizes classic work on the evolution of individual binding sites, and Section III focuses on more recent efforts to simulate the evolution of entire promoter and enhancer sequences. Part II of this review series discusses evolutionary properties and candidate principles that may shape the evolution of gene regulatory architecture. An illustration of key concepts discussed in both Parts I and II, namely *regulatory architecture*, *function*, *grammar* and *code*, is provided by Fig. 1.

## I. GENOTYPE-PHENOTYPE MAPS FOR REGULATORY SEQUENCES

The basic building blocks of GP maps for regulatory sequence evolution comprise the physical mechanisms of regulatory protein-DNA sequence recognition, which are well understood [15,16]. Transcription factors and other regulatory proteins typically recognize $\ell = 6-20$ basepair motifs in the DNA, with shorter recognition lengths surprisingly reported in metazoans [17]. As an example, Fig. 2A–D shows commonly used representations of the binding preferences of the *Escherichia coli* RNAP-σ70 complex for DNA sequences. While these models are statistical and not mechanistic in nature, they have provided an increasingly predictive account of binding, with the latest generation of interpretable [5,18] as well as "black-box" [19] models (Fig. 2E–G) able to predict a substantial fraction, sometimes $\gtrsim 80\%$, of variance in constitutive expression from either completely random or designer DNA sequence libraries - in any case, libraries at a large mutational distance from wildtype promoters - thereby yielding *bona fide* global GP maps for constitutive expression.

To account for regulated - as opposed to constitutive - expression, two extensions to the standard GP map paradigm are necessary. First, the regulatory phenotype can no longer be a single scalar gene expression value [22]: regulation necessarily implies that expression is conditional on cell type or time, extracellular signals, or other causative factors that manifest as changes in intranuclear TF activities or co-factor concentrations. These changes directly impact expression at the CRE locus through differential binding of TFs, and in the fortunate case of GP maps for gene regulation, much is known quantitatively about the relevant TF identities and their levels. From the evolutionary perspective, all these effects are collectively grouped together under the heading of the environment, forming the last leg of the genotype-phenotype-environment triad. The dependence of gene expression level on the environment (i.e., on the environment-specific combinations of TFs concentrations that encode the environment) is the "regulatory function" (Fig. 1B). Different regulatory sequences can implement different regulatory functions, and any putative fitness function must score how well gene expression matches the required level across multiple such "environments".

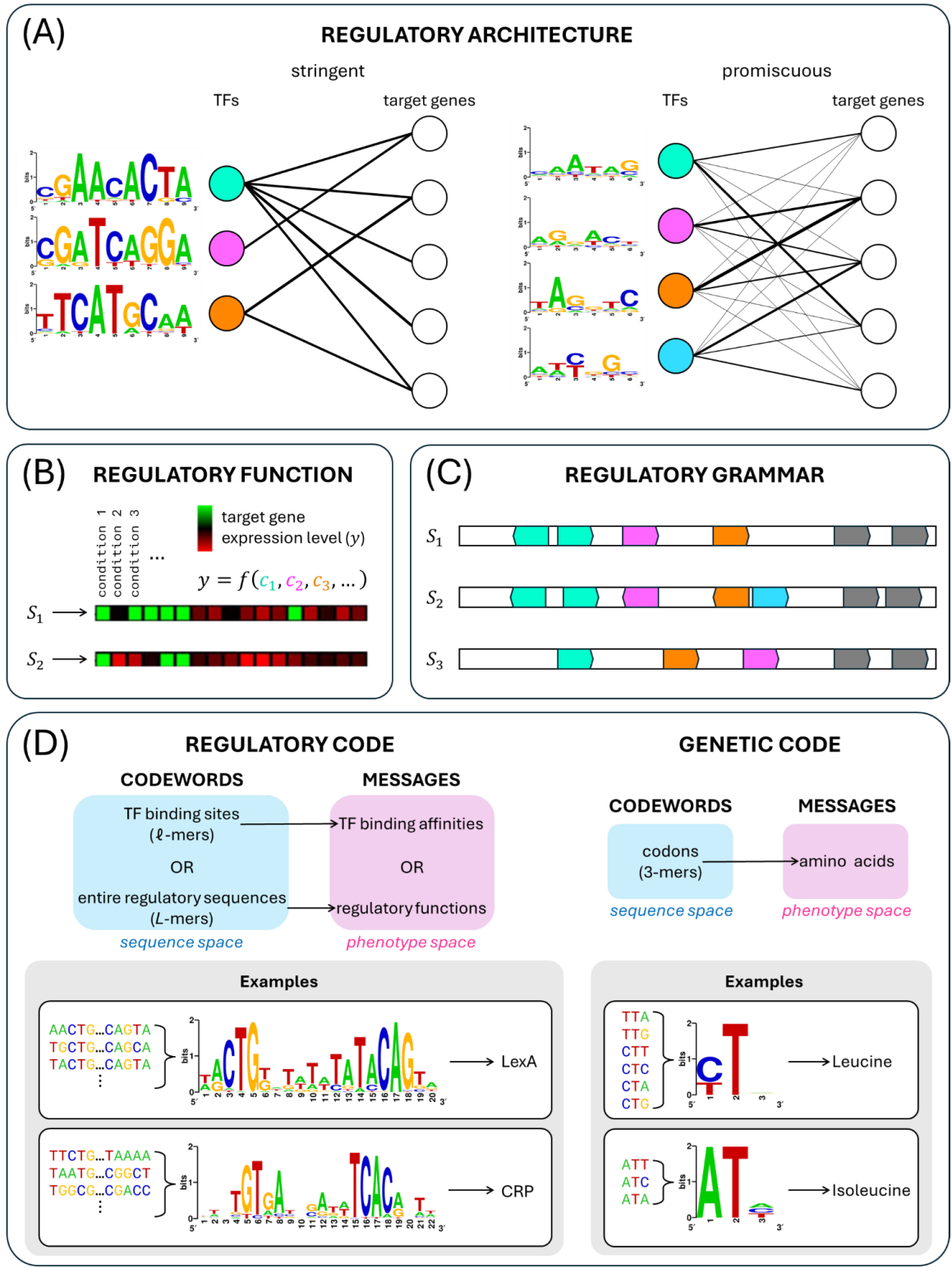

**Figure 1. Illustration of key concepts. (A) Regulatory architecture** refers to the large-scale organization of gene regulation: which genes are regulated by which TFs, whether regulation is layered or hierarchical, and whether interactions are additive or combinatorial. This notion is widely used in genetics, genomics, and systems biology. As an example, panel (A) compares two architectures: on the left, the targets of each TF are sharply distinguished by the non-targets (presence vs absence of a link), through motifs with high information content, with a very uneven number of targets per TF; on the right, the same target genes are instead regulated by a higher number of more promiscuous TFs, with the strength of each potential TF-target interaction being on a continuum (links' width), while the load of regulatory tasks is more homogeneously distributed across TFs. The contrasted features recapitulate some well-known differences between prokaryotic and eukaryotic regulatory architectures. **(B) Regulatory function** is the quantitative mapping from regulatory inputs (encoded by TFs' concentrations) to gene expression output. Any given regulatory sequence implements a regulatory function that makes the expression of the cognate target gene context-dependent (i.e., a function of TFs' concentrations). As input-output relationships, regulatory functions can be approximated by continuous or discrete (logic) mathematical functions. They are often considered in systems and synthetic biology. Panel (B) compares two distinct regulatory functions implemented by two alternative regulatory sequences ($S_1$ and $S_2$). Each condition reported in the expression profile entails different concentrations of the TFs ($c_1, c_2, c_3, ...$), leading to a different expression level for the target gene. **(C) Regulatory grammar** describes the rules by which the number, identity, spacing, orientation, and interactions of TF binding sites determine the regulatory function. Grammar operates at an intermediate level of abstraction: binding sites are treated as functional units (akin to words), and emphasis is placed on how their combinatorial integration shapes gene expression. In this sense, grammar explains how regulatory sequences that contain similar or even identical "vocabularies" of binding sites – like those illustrated in panel (C) – can give rise to different regulatory functions (illustrated in panel (B) for $S_1$ and $S_2$) when sites are arranged differently. This notion is common in genetics, evolutionary and developmental biology. **(D) Regulatory code** is a map in which each possible DNA sequence is associated with a regulatory phenotype. This notion can be applied to cases in which the sequences are all the $\ell$-mers, and the phenotypes are their corresponding binding affinities for a given TF with an $\ell$-bp motif. Alternatively, it can be applied to cases in which the sequences are entire regulatory sequences (of $L$ bp), and the phenotypes are the regulatory functions they implement. Unlike grammar or function, the notion of (regulatory) code is inherently global and concerns the collective assignment of (regulatory) *codewords* (i.e., DNA sequences) to (regulatory) *messages* (i.e., regulatory phenotypes). Panel (D) compares genetic and regulatory codes and provides examples of encoded messages for both. The *code* paradigm is extensively used for the genetic code (hence the term "codon"), but often with minimal application of information theory. For regulatory codes, information-theoretic approaches are standard, but the sets of synonymous sequences are called "motifs" and are seldom formally treated as the codewords of a code.

Second, regulation by multiple TFs requires understanding how the larger regulatory sequence, typically ~200 bp for prokaryotic promoters or eukaryotic enhancers, affects gene expression. The essential simplification here is the approximately "convolutional nature" of the transcriptional regulatory code: TFs still recognize and bind short motifs within this longer sequence in a well-understood fashion, and our job is to devise a mathematical function, known as "regulatory grammar" (Fig. 1C) that describes how multiple such binding sites for various TFs, with different spacing, orientation and cooperativity within the CRE, integrate into multi-valued regulatory phenotypes. While the convolutional nature of the regulation may appear trivial, it represents a drastic simplification compared to handling an entirely unstructured GP map of dimension $4^L$ for L~200, which would be hopelessly out of reach. It also suggests a concrete experimental strategy for dissecting regulatory grammar: by randomly shuffling the positions, orientations and combinations of predefined binding sites and measuring the resulting expression profiles, one can empirically map how spatial organization integrates into regulatory function, as demonstrated by synthetic promoter shuffling approaches [23].

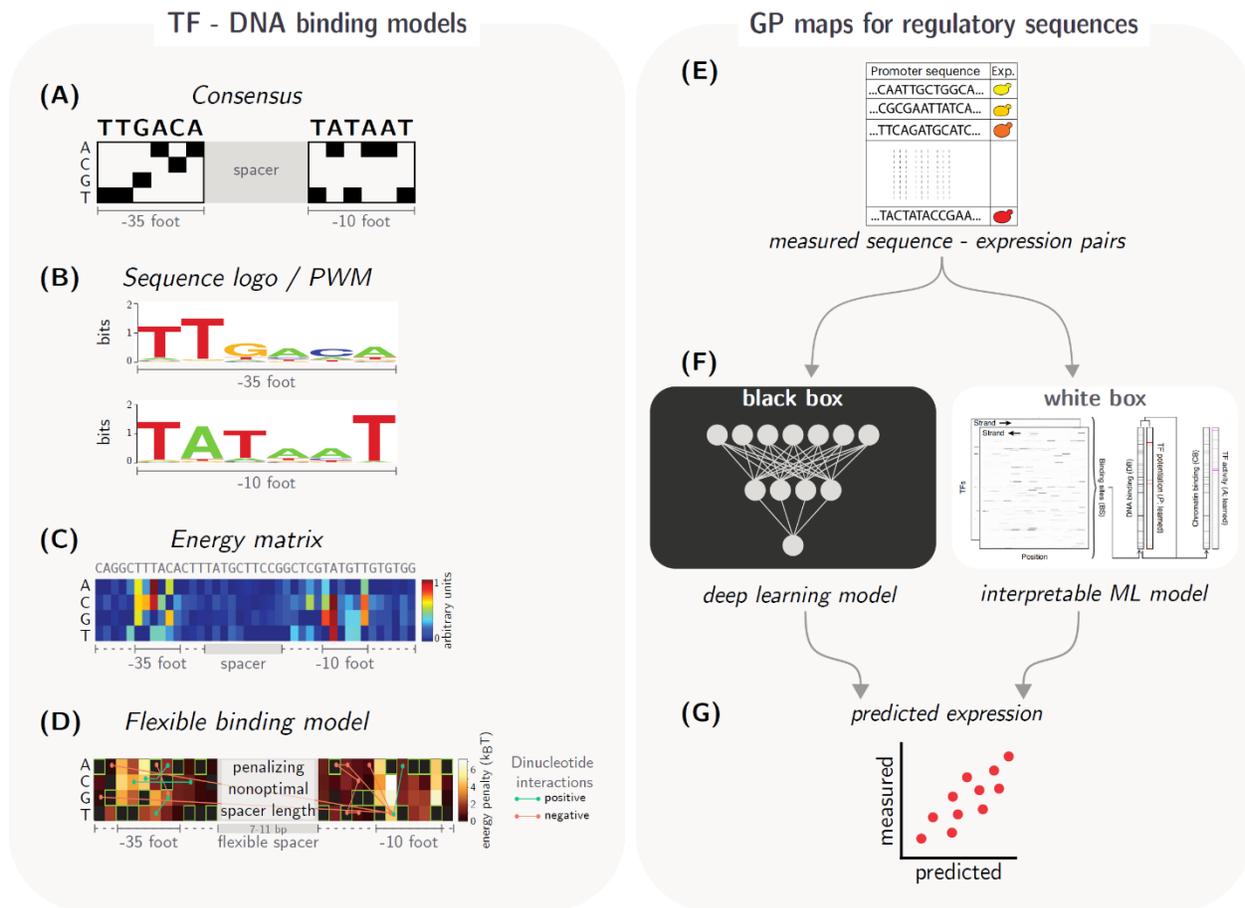

**Figure 2**. **Models that map regulatory sequence to function. (A-D)** Models of TF-DNA binding. In a common thermodynamic framework for gene regulation [20], stronger binding – and thus higher average occupancy of the bacterial promoter by the RNAP – leads to higher constitutive gene expression. **(A)** "Consensus sequence" is a simple summary of RNAP-$\sigma^{70}$ preference to bind two specific sequence hexamers (two "feet") where the DNA is contacted at positions −10 and −35 relative to the transcription start site, separated by a canonical 17bp spacer. The matrix immediately below the sequence shows a simplified "mismatch energy model" often used in theoretical studies: any deviation from the consensus sequence at any position contributes a fixed additive penalty to RNAP-DNA binding energy of $\epsilon \sim 1 - 3 \; k_BT$, set by the scale of hydrogen bonds that underpin molecular recognition. **(B)** Representation of the two feet as "sequence logos", created from a curated and verified list of RNAP binding sites, expresses differential specificity at each binding location for each base. **(C)** Energy matrix representation for RNAP-DNA interaction, which can be calibrated to absolute energy scale, inferred from a massively parallel assay (reproduced from [21]). **(D)** An extended biophysical model that includes a differential energy penalty for spacers between the two feet, and deviations from the additive model in terms of di-nucleotide interactions (reproduced from [5]). **(E-G)** Use of massively parallel assays to infer GP maps for regulatory sequence. Typically, tens of thousands or more mutated versions of a CRE drive a fluorescent reporter (E), and such genotype-expression pairs serve as training samples for fully expressive ("black box") deep learning models (F left), or more interpretable, mechanistically-inspired models fit with modern machine learning tools (F right, reproduced from [6]). The models are tested on withheld data or even mutational libraries with statistical properties that systematically differ from training samples (G).

In prokaryotes, such integration is based on the so-called "thermodynamic models" [20], which represent an extensively tested and trusted paradigm [21,24]. In this paradigm, the thermodynamic equilibrium occupancy of all regulatory factors on the promoter stabilizes or destabilizes the binding of RNAP, whose occupancy monotonically maps into gene expression. Statistical physics provides the mathematical machinery to compute various occupancies given TF concentrations, and to systematically account for known complications such as binding cooperativity, steric occlusion between factors, DNA looping, etc. While some measurements challenge certain thermodynamic models' assumptions [25], these models nevertheless remain a strong, identifiable, and mechanistically grounded baseline GP map for prokaryotic gene regulation [26].

The situation is more complicated in eukaryotes, especially in metazoans. Much is known about how their promoters activate and how signals are integrated across CREs [27,28]. Past groundbreaking efforts to understand eukaryotic regulatory grammar have been successful, especially in yeast [29] and in the context of developmental enhancers [28,30–32], boosted recently by massively parallel experiments and deep learning [6,33,34]. Despite this impressive progress, much remains unknown [35,36]. For example, a definite explanation of why eukaryotic gene regulation utilizes short TF binding sites that individually cannot confer sufficient specificity is still missing; this is called the "specificity paradox" [37]. Similarly unexplained is the functional role of weak, low-affinity binding sites [38]. On the mechanistic side, key to understanding the regulatory grammar and thus the global regulatory GP map is the additivity vs cooperativity (or synergism) of TF binding, a subject of intense theoretical and experimental research [39–41]. Many additional, well-documented sequence-dependent mechanisms implicated in eukaryotic gene regulation (such as chromatin landscapes and accessibility, histone modifications, nucleosome positioning, methylation, loop extrusion and insulation, or non-equilibrium regulatory processes, etc.) are often studied in isolation and remain to be mathematically integrated into an overarching and predictive GP map for eukaryotic gene regulation. Current state-of-the-art either assumes that the effects of these mechanisms, as relevant for the evolutionary outcomes, will be successfully and automatically absorbed by expressive statistical models trained on rich datasets [42], or coarse-grains many such putative mechanisms into an effective mathematical approximation, e.g., by composing the GP map from successive yet understandable linear-nonlinear transformations [43], much like in computational neuroscience or neural networks. These approaches, schematized in Fig. 2E–G, are not exclusive and pursue, from complementary directions, the same question: Which mechanistic features of gene regulation substantially affect CRE evolutionary trajectories, and which can be safely ignored? Much exciting work remains to be done on this front.

## II.    EVOLUTION OF INDIVIDUAL TRANSCRIPTION FACTOR BINDING SITES

Models for single TF binding presented in Fig. 2A–D allow us to study the emergence of TF binding sites *de novo* from a random (or some alternative) starting sequence ensemble, as well as their maintenance and turnover. The simplified mismatch model in Fig. 2A enables many analytical treatments, which typically agree well with simulations based on more detailed but analytically intractable models (Fig. 2B–D). The simplest setup assumes directional selection for binding of a TF, either within a restricted sequence window of size L=ℓ bp (where ℓ is the length of the TF's motif) or larger, the size of the entire CRE (L≫ℓ). In both cases, the sequence is typically short enough to neglect recombination on the timescales of interest. Two basic GP map features are key to the resulting evolutionary dynamics: (i) effects of individual basepairs first combine linearly (via the mismatch model or the energy matrix) into TF binding energy; this step captures the huge degeneracy of sequence space without intrinsic higher-order (epistatic) interactions; (ii) the resulting energy maps into binding probability, and thus fitness, via a sigmoid "binding" nonlinearity, as dictated by the thermodynamics of TF-DNA interactions. This nonlinearity induces large neutral plateaus (where the nonlinearity is flat) in sequence space.

Several approaches can be used to assess the extent to which selection can give rise to *de novo* TF binding sites and maintain them against the entropic forces imposed by mutation and drift. Explicitly modeling mutation rates leads to a mutation-selection balance, where deleterious mutations continually erode binding affinity and selection counteracts this loss. Even in the infinite-population limit (no drift), selection cannot fully concentrate probability mass on the consensus sequence because mutations continually redistribute it across sequence space [44]. Alternatively, one can work in the strong-selection, weak-mutation limit, often analyzed using the fixed-states approximation: the population is assumed fixed for a single genotype, and evolves via mutations that either fix or are lost according to Kimura's fixation probability. Here, mutation acts merely as a generator of variation, while the dominant entropic force opposing fitness maximization is genetic drift. The resulting evolutionary steady state reflects the drift-selection balance [45]. These entropic forces map onto the energy-entropy trade-off of statistical physics [46,47] and explain on evolutionary grounds why mismatches between the consensus and functional binding sites are to be expected. When applied to de novo emergence of binding sites, these models predict rapid evolution via point mutations for very short sites [48], but exponentially longer times as site length ℓ increases, which is difficult to reconcile with comparative genomics evidence for fast turnover at realistic ℓ, unless selection strength is exceedingly strong [49]. Slow dynamics arise because random initial sequences are far from consensus, confining adaptation to a random walk over a vast, nearly flat fitness landscape with a vanishing selection gradient (due to the binding nonlinearity). This limitation is not easily circumvented, and active research is directed towards mechanisms that could accelerate adaptation (such as noise [50]; sequence changes beyond point mutations [51,52]; promiscuity-inducing mutations that decrease TFs' specificity [53]; and recognition beyond rigid motifs, such as flexible spacers between short motifs [5,54], Fig. 2D).

A valuable counterpart to the theoretically-driven studies are data-driven approaches that learn selection pressures or fitness landscape parameters from genomic data, using physical models (e.g., thermodynamic models of TF-DNA interaction) to make such inferences regularized and well-defined. Effective fitness landscapes and selection pressures for RNAP and TF binding sites have been inferred from bioinformatic analyses in prokaryotes and yeast [55,56]. Beyond the direct inference of the GP map, clever experimental designs, e.g., using yeast hybrids, allow for the dissection of specific mechanisms and evolutionary forces important for the evolution of gene regulation [57].

### III. EVOLUTION OF ENTIRE REGULATORY SEQUENCES

Simulating the evolution of entire CREs is challenging: not so much due to the technicalities or the raw simulation runtime – especially given today's computing power – but mainly due to the many structural and parametric assumptions one needs to make to instantiate a quantitative and global GP map, especially in metazoans, as well as the difficulty of making sense of simulated outcomes. Despite these difficulties, some (though not many) such simulations have been attempted, as reviewed below, opening up opportunities for larger-scale exploration in the near future.

Early attempts to scale up from individual TF binding site evolution towards entire CREs focused on overlapping and competing TF binding sites and their functional importance within the CREs. The thermodynamic model was used to derive a GP map with overlapping and occluding sites, allowing the authors to draw relatively general conclusions relevant to pro- or eukaryotes, while simplifying some aspects of the actual evolutionary process [58]. Further generic conclusions were pursued in a stylized biophysical model to ask about TF binding site turnover in a CRE while conserving regulatory function, focusing on selection strength and fitness effects of mutations over evolutionary time; importantly, this work was early to emphasize the importance of non-cognate binding that could give rise to deleterious regulatory crosstalk [59,60]. Another qualitatively new and generic mechanism available when considering the evolution of entire CREs is the summation of multiple, possibly overlapping, binding-site effects (as empirically demonstrated [38]), in particular when enabled by short tandem sequence repeats [61].

Moving from generic results to system-specific and quantitative work, the evolution of constitutive bacterial promoters was experimentally studied and computationally simulated within the scope of an improved "flexible RNAP binding model," inferred via a massive mutagenesis essay (Fig. 2D) [5]. GP maps derived from this model drastically increased the fraction of random sequences that drove significant expression, as reported previously [62], as well as the number of non-expressing sequences that are one mutational step away from expressing; such sequence predictions were successfully experimentally verified. Importantly, the flexible model predicts order-of-magnitude increases in the rate of *de novo* constitutive promoter evolution relative to previous state-of-the-art models, rationalizing the existence of specific mechanisms that might appear unimportant without the evolutionary perspective. These findings served as a stepping stone towards understanding the evolution of not only constitutive, but also regulated bacterial promoters [22,63].

Moving beyond prokaryotes, GP maps based on statistical models inferred from massively parallel assays for yeast regulatory sequences [6] enabled intriguing evolutionary interrogation of how directional or stabilizing selection would act on CREs, even when the evolutionary dynamics employed were not entirely realistic [64]. Nevertheless, the key questions relevant for any subsequent theory of regulatory sequence evolution (see Part II of this review series) were addressed, including: a systems-level approach, diminishing returns epistasis as a consequence of a realistic GP map, regulatory phenotypes that are high-dimensional due to conditioning on the environment, and the importance of robustness and evolvability in understanding the inferred map, suggesting that these properties may have emerged or been selected for over the long epochs of evolutionary time.

In metazoans, the simulation of regulatory sequence evolution for developmental enhancers has an even longer history. Theoretical work demonstrated that *de novo* emergence of transcription factor binding sites from truly random initial CRE sequence is prohibitively slow [49], but left open the path to such emergence from non-random initial sequence that could contain the so-called "pre-sites," whose importance has been suggested earlier [65]. Here, too, the possible importance of overlapping TF binding sites [66], as well as clusters of (potentially weaker) multiple binding sites [67] has been highlighted. There have been several attempts to integrate the known regulatory phenomenology and modeling of CREs involved in early *Drosophila* patterning to simulate regulatory sequence evolution [68], including focusing on the timescale for such evolutionary adaptations [69]. While very promising, such work on the sequence level remains to be integrated with a theory of how individual gene expression phenotypes and patterns contribute to organismal fitness, as suggested by normative theories [70]: in the opinion of the authors, this is one of the key open **Challenges**, highlighted in Part II of this review series.

On the experimental front, the direction that might, perhaps surprisingly, interact most strongly with theory is the use of truly random massive mutational libraries for regulatory sequences. By "truly random" we mean constructs that do not mutagenize wild-type promoters or enhancers, but explore the entire sequence space randomly – in the extreme, by generating regulatory sequences with background A, C, T, G probabilities, with no correlation between positions. A standard textbook viewpoint holds that a completely random sequence is typically inert, resulting in no downstream gene expression, because the probability of containing a functional sequence is small enough to be negligible. Contrary to this expectation, many completely random promoter sequences can drive significant expression in bacteria – around 10% of the randomly generated sequences or more, depending on sequence length and other details – but in any case far larger than that expected based on naive consensus sequence matching [5,62,71]. Related and intriguing results are not limited to prokaryotes. In yeast, utilizing random promoter sequences has also been suggested as a non-intuitive but productive means of GP map exploration [72], with strong indications that this approach extends to multicellular eukaryotes [73], including very exciting recent results in the fruit fly [74]. What gives further credence to the random library approach is also the associated theory work [22,43,75,76].

The empirical and modeling results reviewed here suggest that the information coding for regulatory phenotypes evolves in ways that depend on properties of the genotype-phenotype-fitness map that can be characterized and quantified. Some of these properties are themselves genetically encoded, raising the possibility that not only regulatory sequences, but the regulatory code itself might evolve. This parallels decades of work on the genetic code suggesting that, although extant protein-coding sequences share a (nearly) universal code, the architecture of the code itself might have evolved during pre-LUCA stages of life, favoring encoding strategies that jointly provide mutational robustness and evolvability [77,78]. In Part II of this review series, we discuss how developments at the interface of information theory and evolutionary theory can quantify the action of selection, mutation and drift on regulatory sequences and codes, and outline future directions toward a unifying language for their emergence and organization.

**Acknowledgments**

We thank Nick Barton and Noa Ottilie Borst for essential contributions to this manuscript.

E.M. acknowledges support from the APART-USA fellowship, jointly funded by the Austrian Academy of Sciences (ÖAW) and the Institute of Science and Technology Austria (ISTA).

This study was supported by the European Molecular Biology Laboratory (N.O.B., J.C.); the European Molecular Biology Laboratory Interdisciplinary Postdoc Programme (EIPOD) under the Marie Skłodowska-Curie Actions cofund (S.H.A.).

78.	Rozhoňová H, Martí-Gómez C, McCandlish DM, Payne JL. Robust genetic codes enhance protein evolvability. PLOS Biol. 2024 May 16;22(5):e3002594.